\documentclass[amsmath,amssymb,pra,aps,showpacs,superscriptaddress,twocolumn,notitlepage]{revtex4-2}
\usepackage{amsmath,amsfonts,amssymb,amsthm,graphics,graphicx,epsfig,bbm}
\usepackage[colorlinks=true,citecolor=blue,linkcolor=blue,urlcolor=blue]{hyperref}
\usepackage[usenames]{color}
\usepackage{graphicx}
\usepackage{subfigure}
\usepackage{amsmath}
\usepackage{epsfig}
\usepackage{dcolumn}
\usepackage{bm}
\usepackage{color}
\usepackage{times}
\usepackage{epstopdf}
\usepackage{amssymb}
\usepackage{amstext}
\usepackage{latexsym}
\usepackage{hyperref}
\usepackage{amsfonts}
\usepackage{psfrag}
\usepackage{soul,xcolor}
\usepackage[normalem]{ulem}
\usepackage{dsfont}
\usepackage{adjustbox}

\renewcommand{\Re}{\mathrm{Re}}
\renewcommand{\Im}{\mathrm{Im}}
\newcommand{\ket}[1]{\vert #1 \rangle}

\newcommand{\ketbra}[2]{\vert #1 \rangle \langle #2 \vert}

\begin{document}
	\title{Shortcuts to adiabaticity  in superconducting circuits for fast multi-partite state generation}
	
	\author{Francisco Andrés C\'ardenas-L\'opez}
	\email{f.cardeans.lopez@fz-juelich.de}
	\affiliation{International Center of Quantum Artificial Intelligence for Science and Technology (QuArtist)\\
		and Physics Department, Shanghai University, 200444 Shanghai, China}
	\affiliation{Forschungszentrum J\"ulich GmbH, Peter Gr\"unberg Institute, Quantum Control (PGI-8), 52425 J\"ulich, Germany}

	\author{Juan Carlos Retamal}
	\affiliation{Departamento de F\'isica, Universidad de Santiago de Chile (USACH), Avenida V\'ictor Jara 3493, 9170124, Santiago, Chile}
	\affiliation{Center for the Development of Nanoscience and Nanotechnology, CEDENNA, Estaci\'on Central, 9170124, Santiago, Chile}
	
	\author{Xi Chen}
	\email{xi.chen@ehu.eus}
	\affiliation{Department of Physical Chemistry, University of the Basque Country UPV/EHU, Apartado 644, E-48080 Bilbao, Spain}
	\affiliation{EHU Quantum Center, University of the Basque Country UPV/EHU, 48940 Leioa, Spain}
	
	\maketitle
\onecolumngrid 
\section*{abstract}
Shortcuts to adiabaticity provides a flexible method to accelerate and improve a quantum control task beyond adiabatic criteria. However, their application to fast generation of multi-partite quantum gates is still not optimized. Here we propose the reverse-engineering approach to design the longitudinal coupling between a set of qubits coupled to several field modes, for  achieving a fast generation of multi-partite quantum gates in photonic or qubit-based architecture. We show that the enhancing generation time is at the nanosecond scale that does not scale with the number of system components. In addition, our protocol does not suffer noticeable detrimental effects due to the dissipative dynamics. Finally, the possible implementation is discussed with the state-of-the-art circuit quantum electrodynamics architecture. \\
\twocolumngrid
	\section*{Introduction}
	It has been scrutinized that  entanglement~\cite{RevModPhys.81.865,Adesso2016} in multi-partite quantum systems, plays a crucial role in quantum technologies applications such as quantum information processing~\cite{PhysRevLett.85.2392,Nature416238a,nature35005001}, quantum computation~\cite{Feynmann1981,nature35005001}, and quantum simulation~\cite{Ekert1998}, respectively. Entanglement as a quantum resource can lead  to a speed-up in the running time of quantum algorithms~\cite{DJalgorithm,Salgorithm,Galgoritm,Algorithmreview}. Furthermore, the entanglement characterization in  many-body systems may provide helpful information concerning whether it is possible to find underlying features of the low-lying energy spectrum with accurate numerical methods~\cite{Rev.Mod.Phys.77.259,Phys.Rev.B.48.10345}. Likewise, entanglement discontinuity is an excellent indicator to characterize phase transitions on quantum systems~\cite{PhysRevA.66.032110,N416608a,PhysRevLett.93.086402,PhysRevLett.105.095702}. Besides, in the context of quantum metrology and sensing, using entangled states has achieved quantum-enhanced precision measurements near to the shoot-noise limit~\cite{PhysRevLett.96.010401,PhysRevLett.101.040403,nature08988}, among other applications.
	
	A fundamental condition to generate such multi-partite entangled states relies on the capability of our quantum platform to access a set of interactions or controlled quantum gates that communicate all the system components (all-to-all). Feasibly controllable trapped-ion platform~\cite{Tionsreview} gives a step forward in this direction with the implementation of the so-called S\o rensen-M\o lmer quantum gate~\cite{PhysRevLett.82.1971,PhysRevLett.82.1835,PhysRevLett.86.3907}, where external lasers applied to a confined array of alkali atoms permit access to the red and blue sidebands so that the effective interaction between the ions is of the form $\sigma_{\ell}^{x}\sigma^{x}_{\ell'}$~\cite{PhysRevLett.91.157901,PhysRevA.71.062309}. 

	On the other hand, it is possible to engineer similar interactions in superconducting quantum circuit and circuit quantum electrodynamics (cQED) ~\cite{Devoret2005book,You2005,Clarke2008,Wendin2005,Devoret2013,Kockum2019,Krantz2019,Kjaergaard2020,Martinis2020,Phys.Rev.A.69.062320,Nature.431.162,Nature.431.159,Nature.451.664,arXiv.2005.12667,Blais2020}, where Josephson junction-based electrical circuits having discrete energy spectrum mimics artificial atoms~\cite{Bouchiat1998,Nakamura1999,Koch2007,Schreier2008,Barends2013,Martinis2002,Steffen2002,Ansmann2009,Orlando1999,Mooij1999,You2007,PhysRevLett.105.100502,ncomms12964,Shcherbakova2015}, whereas quantized field modes corresponds to either LC resonators, coplanar waveguide or stripline resonator~\cite{Girvin2011,Nori3,Itoh1974,Goppl2008,Gely2017}. In this architecture, it is possible to engineer multi-qubit interactions by coupling all of them to a common resonator bus such that in the dispersive regime, we obtain a like-S\o rensen-M\o lmer interaction~\cite{PhysRevB.74.104503,PhysRevB.81.104524,PhysRevA.95.013845,Scirep.9.1380} as a result of a second-order interaction which in principle are slower than the resonant case make them fragile against the unavoidable action with environment~\cite{PhysRevA.65.052327,PhysRevB.66.193306,PhysRevLett.93.230501,PhysRevLett.107.090501}.

	Enhanced performance of these protocols may require access to higher coupling strength values where the system operates in the so-called ultrastrong or deep strong coupling regime~\cite{Nori2,FornDiaz2019,Casanova2010,Yoshihara2017}. In such regimes, however, the high hybridization of the energy levels makes it difficult to distinguish between the light and matter degree of freedom. An alternative approach relies on engineering either a pulse sequence or the coupling strength following the ubiquitous methods of shortcut to adiabaticity (STA)~\cite{PhysRevLett.104.063002,RevModPhys.91.045001} that allow us to control a quantum system to accelerate an adiabatic evolution overcoming preparation errors and minimizing the action of the environment~\cite{RevModPhys.91.045001}. STA has received renewed interest in the context of cQED since it has been generalized to open quantum system~\cite{PhysRevLett.122.250402,Quantum} permitting to design a counter-diabatic and optimal pulses to speed up a dissipative evolution~\cite{AnNC,Panchito}. 
	Motivated by this, we propose a reverse-engineering method to accelerate the generation of multi-partite entangled states. We design a modulated longitudinal coupling strength that accelerate the generation of multi-partite photnonic/qubit states within the nanosecond scale that does not scale with the number of systems (field modes and qubits). Because of the short time, we observe no detrimental effect produced by the action of the environment. Finally, we propose the possible implementation in cQED architecture.
	
	\section*{Results}
	\subsection*{Longitudinal Interaction}
	
	\label{TheModel}
	We start describing the fundamental ingredient in generating multi-partite entangled states. To do so, let us consider a set of $N$ two-levels systems of frequency $\Omega_{n}$ coupled to $M$ quantized field modes of frequency $\omega_{m}$ through time-dependent longitudinal coupling $g_{n}^{m}(t)$ governed by the following Hamiltonian ($\hbar \equiv 1$) 
	\begin{eqnarray}
		\label{model} 
		\mathcal{H}_{n}^{m} = \sum_{n}\frac{\Omega_{n}}{2}\sigma^{x}_{n} &+& \sum_{m}\omega_{m}a_{m}^{\dag}a_{m}  \nonumber \\ 
		&+& \sum_{n,m} g^{m}_{n}(t)\sigma^{x}_{n}(a_{m}^{\dag} + a_{m}),~~~~
	\end{eqnarray}
	where $\sigma^{x}_{n}$ corresponds to the $x$-component Pauli matrix characterizing the $n$-th two-level system, and $a_{m}$ ($a_{m}^{\dag}$) stands for the annihilation (creation) bosonic operator of each field mode, respectively. In cQED systems, we can engineer the longitudinal interaction in an artificial atom coupled to a resonator sharing the common external magnetic flux~\cite{PhysRevLett.115.203601,PhysRevA.95.052333}, or couple both subsystems through a superconducting quantum interference device  (SQUID)~\cite{PhysRevA.80.032109,PhysRevB.91.094517}.

	On the other hand, the structure of the longitudinal coupling between two-level systems with a quantized field mode is suitable for implementing reversed engineering protocols since both interaction and free energy term commutes. Such relation allows us to find an exact adiabatic transformation that accelerates quantum processes such as the qubit readout~\cite{Panchito} or implementing faster two-qubit gates~\cite{Panchito2}, and now we will use it to accelerate the generation of multi-partite entangled states of either photonic or qubit states, respectively.
	
	\par
	
	To generate entangled photonic states, we assume that the field modes couple to a single two-level system, i.e., $N=1$. In such a case, the Hamiltonian reads
	\begin{eqnarray}
		\label{single_qubit}
		\mathcal{H}_{1}^{m} = \frac{\Omega}{2}\sigma^{x} + \sum_{m}\omega_{m}a_{m}^{\dag}a_{m}+\sum_{m} g_{1}^{m}(t)\sigma^{x}(a_{m}^{\dag} + a_{m}).~~~
	\end{eqnarray}
	The Hamiltonian dynamics $\mathcal{H}_{1}^{m}$, in the interaction picture, corresponds to a state-dependent cavity drive represented by the following time-evolution operator  
	\begin{eqnarray}
		\label{single_qubit_int}
		U^{m}_{1}(t)=\prod_{m}\hat{\mathcal{D}}_{m}(\alpha_{m}(t)\sigma^{x}),
	\end{eqnarray}
	where $\hat{\mathcal{D}}_{m}(\sigma^{x}\alpha_{m}(t))=\exp\big(\sigma^{x}(\alpha_{m}(t) a_{m}^{\dag}+\alpha_{m}^{*}(t) a_{m}))$ is the displacement operator of the $m$th field mode with $\alpha_{m}(t)=-i\int_{0}^{t}g_{m}(s)e^{i\omega_{m}s}ds$ as the cavity displacement. For a time-independent coupling strength, we obtain that at time $T_{k}=(2k+1)\pi/\omega_{m}$ the field mode reaches its maximum displacement $\alpha_{{\rm{max}}}=\pm2g_{1}^{m}/\omega_{m}$ that depends on the qubit state. Thus, for the system prepared in the initial state $\ket{\Psi(0)}=\ket{g}\bigotimes_{m=1}^{M}\ket{0}_{m}$ ($\ket{g}$ is the ground state of the two-level system, and $\ket{0}_{m}$ is the vacuum state of the $m$th field mode) the system evolves to the so-called Greenberger–Horne–Zeilinger cat state~\cite{PhysRevA.72.022320,PhysRevLett.77.4887,nature13436,science.1243289}
	\begin{eqnarray}
		\label{GHZ_cat0}
		\ket{\Psi(T)}=\frac{1}{\sqrt{2}}\bigg[\ket{e}\bigotimes_{m=1}^{M}\ket{\alpha_{{\rm{max}}}}_{m}+\ket{g}\bigotimes_{m=1}^{M}\ket{-\alpha_{{\rm{max}}}}_{m}\bigg],~~~
	\end{eqnarray}
	by rotating the qubit state along the $y$ axis, we obtain
	\begin{eqnarray}
		\label{GHZ_cat}
		\ket{\Psi}=\frac{1}{\sqrt{2}}\bigg[\ket{e}\bigotimes_{m=1}^{M}\ket{\alpha_{+}}_{m}+\ket{g}\bigotimes_{m=1}^{M}\ket{\alpha_{-}}_{m}\bigg],
	\end{eqnarray}
	where $\ket{\alpha_{\pm}}_{m}=(\ket{\alpha_{{\rm{max}}}}_{m}\pm \ket{-\alpha_{{\rm{max}}}}_{m})/\sqrt{2}$ is the even/odd coherent state superposition~\cite{Scully_quantum_optics_book}. Thus we are lead with a multi-partite hybrid light-matter state embedding maximal entanglement, not being affected by local operations and classical communication (LOCC)~\cite{Nielsen_chuang}.\par
	
	On the other hand, it is possible to generate multi-partite entangled states of qubits when we consider that $N$ two-level systems are coupled to a single field mode described by the Hamiltonian
	\begin{eqnarray}
		\label{Hamiltonian_GHZ_qubit}
		\mathcal{H}_{n}^{1} = \omega a^{\dag}a + \sum_{n}\frac{\Omega_{n}}{2}\sigma^{x}_{n} +\sum_{n} g^{1}_{n}(t)\sigma^{x}_{n}(a^{\dag} + a).
	\end{eqnarray}
	In the interaction picture,  the time-evolution operator of $\mathcal{H}_{n}^{1}$ can be written in a factorized form using the Baker-Campbell-Hausdorff (BCH) formula~\cite{BHf}
	\begin{eqnarray}
		\label{U_GHZ_qubit_int_picture}
		U_{n}^{1}(t) = \prod_{n}e^{-iA_{n}\sigma^{x}_{x}a}\prod_{n}e^{-iA_{n}^{*}\sigma^{x}_{x}a^{\dag}}\prod_{n, n'}e^{-iB_{nn'}\sigma^{x}_{n}\sigma^{x}_{n'}}.
	\end{eqnarray}
	The coefficients $A_{n}(t)$ ($A_{n}^{*}(t)$) and $B_{nn'}(t)$ are obtained calculating the Schr\"odinger equation for the time-evolution operator $i~\dot{U}_{n}^{1}(t)=\mathcal{H}_{n}^{1}U_{n}^{1}(t)$ (we refer to the reader to the Supplementary Note 1 for the detailed calculation), we obtain that these coefficients must satisfy the following differential equations
	\begin{eqnarray}
		\label{equation_of_motion}
		\frac{dA_{n}(t)}{dt}=g^{1}_{n}(t)e^{-i\omega t},\quad \frac{dB_{nn'}(t)}{dt}=iA_{n'}(t)\frac{dA_{n}^{*}(t)}{dt}.
	\end{eqnarray}
	For time-independent coupling strength $g_{n}^{1}$ with the initial conditions $A_{n}(0)=B_{nn'}=0$, we obtain~\cite{PhysRevB.81.104524}
	\begin{subequations}
		\begin{eqnarray}
			\label{solution_equation_of_motion}
			A_{n}(t)&=&\frac{ig^{1}_{n}}{\omega}(e^{-i\omega t}-1),\\
			B_{nn'}(t)&=&\frac{g^{1}_{n}g^{1}_{n'}}{\omega}[-i(e^{i\omega t}-1)-t].
		\end{eqnarray}
	\end{subequations}
	By choosing $\tau=2\pi/\omega$, the coefficient $A_{n}(\tau)$ vanishes and $B_{nn'}(\tau)\equiv\theta_{nn'}=2\pi g^{1}_{n}g^{1}_{n'}/\omega^{2}$. Consequently, the time evolution operator reduces to 
	\begin{eqnarray}
		\label{SMGate}\nonumber
		U_{n}^{1}(\tau) &=&\prod_{n, n'}\exp\bigg[-i\theta_{nn'}\sigma^{x}_{n}\sigma^{x}_{n'}\bigg]\\
		&=&\prod_{n\neq n'}\bigg[\cos(\theta_{nn'})\mathbb{I}+i\sin(\theta_{nn'})\sigma^{x}_{n}\sigma^{x}_{n'}\bigg],
	\end{eqnarray}
	corresponding to the S\o rensen-M\o lmer quantum gate~\cite{PhysRevLett.82.1971,PhysRevLett.82.1835} (SMG). It is worthy to mention that the SMG has been widely used in the context of quantum simulation to codify fermionic system in a set of coupled two-level system~\cite{fermion1,fermion2}. If we initialize the system in the state $\ket{\Phi(0)}=\bigotimes_{n=1}^{N}\ket{g}_{n}\otimes\ket{0}$, and choose $\theta_{nn'}=\pi/4$, the state evolves to a GHZ state~\cite{PhysRevA.62.062314}
	\begin{eqnarray}
		\label{GHZ_state_qubit}
		\ket{\Phi(\tau)}=\frac{1}{\sqrt{2}}\bigg[\bigotimes_{n=1}^{N}\ket{g}_{n}+e^{i\pi (N+1)/2}\bigotimes_{n=1}^{N}\ket{e}_{n}\bigg]\ket{0},
	\end{eqnarray}
	Notice that in those derivations, the time generation is constrained by the ratios $g_{1}^{m}/\omega_{m}$, and $g_{n}^{1}/\omega_{m}$, respectively. In particular, it is possible to implement the multi-partite gate for photons in the timescale $T=47.61~\rm{(ns)}$, whereas for the multi-partite operation with qubits, the gate time is $T=19~\rm{(ns)}$~\cite{PhysRevB.81.104524}. Thus, to obtain faster time generation, we must achieve larger coupling strength values, meaning that the system will operate in the ultra-strong or deep-strong coupling regime~\cite{Nori2,FornDiaz2019,Casanova2010,Yoshihara2017}. A way to accelerate the generation time without demanding larger coupling strength relies upon the technique of  STA~\cite{RevModPhys.91.045001}, where the modulation of system parameters or the addition of an additional term on the system dynamics leads to a speedup of the quantum processes. In what follows that we will explore the former alternative,  the reverse-engineering method \cite{PhysRevLett.104.063002,Panchito}, to design the coupling strength $g^{n}_{1}$ and $g^{1}_{n}$ such that the gating time is shortened significantly.
	
	\begin{figure}[!t]
		\centering
		\includegraphics[width=1\linewidth]{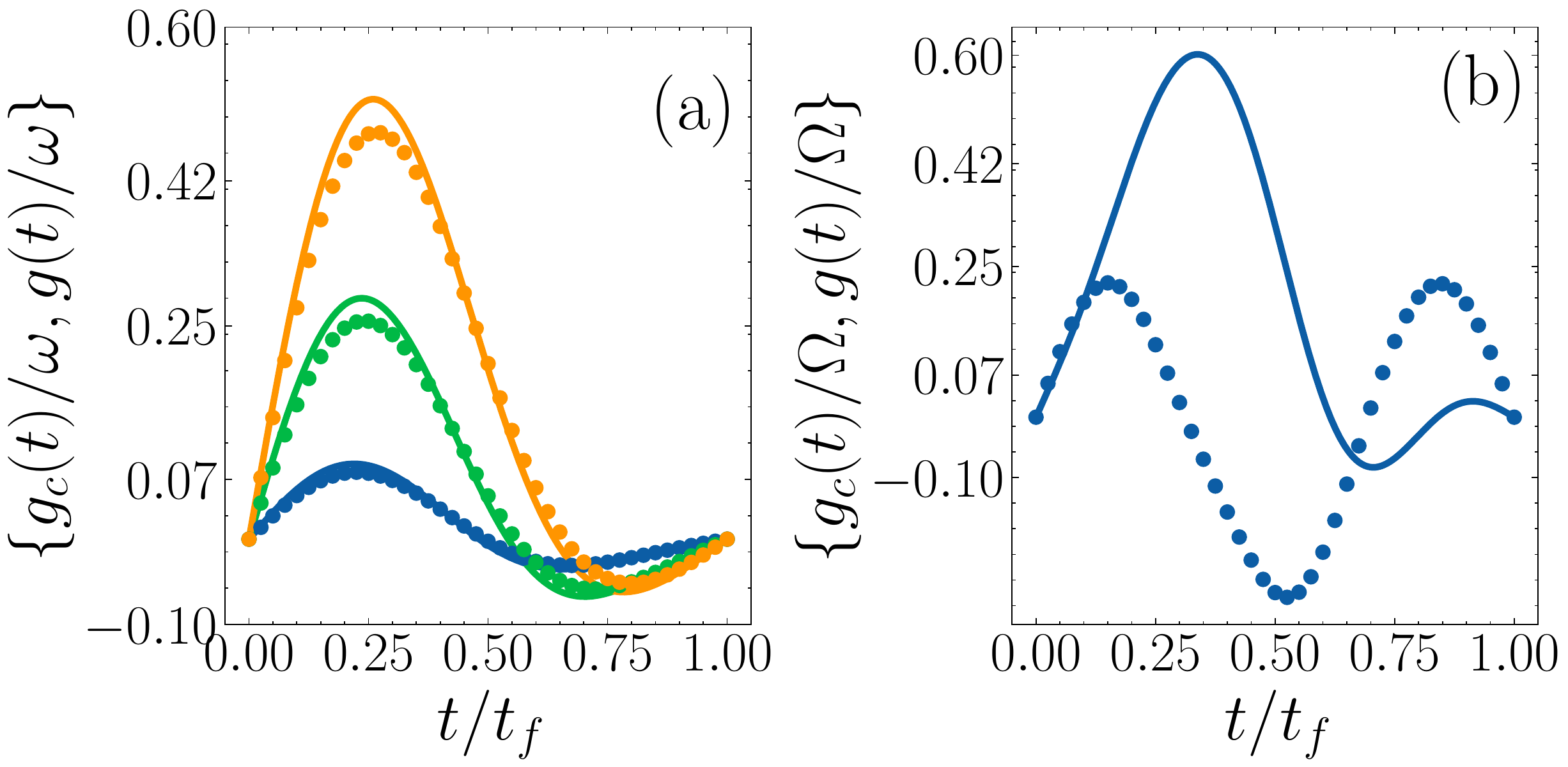}
		\caption{\textbf{Coupling strength modulation}. Reversely engineered coupling strength $g_{c}(t)$ (continuous line) and $g(t)$ (dotted) as a function of the dimensionless time $t/t_{f}$ for (a) the GHZ for photonic systems, blue, orange and green data corresponds to the pulse considering $\alpha(t_{f})=1,3,5$, respectively. Whereas (b) stands for the GHZ in qubits. The numerical simulations are preformed with the following system parameters (a) $\omega/2\pi=6.6~{\rm{GHz}}$, $g/2\pi=21~{\rm{MHz}}$, and (b) $\Omega/2\pi=10~{\rm{GHz}}$, $\omega/2\pi=1~{\rm{GHz}}$, and $g/2\pi=144~{\rm{MHz}}$.}
		\label{Fig_modulation}
	\end{figure}

	\subsection*{Reverse engineering protocol}
	\label{STA}
	Afterwards defining the model, we will discuss how to design the coupling strengths $g^{n}_{1}$ and $g^{1}_{n}$ in the frame of reverse-engineering approach \cite{PhysRevLett.104.063002,Panchito}. We will develop the theory considering a single two-level system ($N=1$) coupled to a single field mode ($M=1$) with coupling strength $g(t)$, the extension to many qubits and field modes is easily derivable. In this case, we propose the solution of the Schr\"odinger equation of the form $\ket{\Psi(x,t)}=\exp{(-it\epsilon)}\mathcal{U}(t)\ket{\varphi(x,t)}$, where $\epsilon=\omega(n+1/2)$ and $\ket{\varphi(x,t)}=\sqrt[4]{m\omega_{r}/\pi}\exp(-m\omega_{r}x^2/2)$ corresponds to the eigenfunction for the uncoupled cavity with Hamiltonian $\omega_{r} a^{\dag}a$ and $\ket{\xi}\equiv\ket{g}$ describes the ground state of the qubit. Additionally, $\mathcal{U}(t)$ stands for a unitary transformation that eliminates the qubit-field mode coupling~\cite{PhysRevA.97.013631,New.J.Phys.15.013029,PhysRevLett.112.150402} 
	\begin{eqnarray}
		\label{unitary}
		\mathcal{U}(t)=e^{i\beta(t)}e^{-i\dot{g}_{c}(t)\sigma^{x}(a^{\dag}+a)/\omega^{2}}e^{-g_{c}(t)\sigma^{x}(a^{\dag}-a)/\omega}.~~
	\end{eqnarray}
	Here, the overdot notation represents time-derivative, and $\beta(t) = -\int_{0}^{t}\mathcal{L}_{g}(s)ds$ corresponds to a phase factor that relates the coupling strength $g(t)$ with the auxiliary variable $g_{c}(t)$ through the classical Lagrangian
	\begin{eqnarray}
		\label{phases_lagrangian}
		\mathcal{L}_{g}(t) &=&\frac{\dot{g}^{2}_{c}(t)}{\omega^{3}} - \frac{g^{2}_{c}(t)}{\omega} - \frac{2g_{c}(t)g(t)}{\omega}.
	\end{eqnarray}
	To guarantee that $\ket{\varphi(x,t)}$ corresponds to the exact solution of the time-dependent Sch\"odinger equation, the classical variable must obey the following equation of motion
	\begin{eqnarray}
		\label{ELE}
		\ddot{g}_{c}(t) + \omega^{2}[g_{c}(t)+g(t)]=0,
	\end{eqnarray}
	which is nothing but the Euler-Lagrange equation from the classical Lagrangian. The auxiliary variable $g_{c}(t)$ needs to satisfy the following boundary conditions to guarantee that at the initial/final time, the auxiliary variable does not participate in the system
	\begin{subequations}
		\begin{eqnarray}
			\label{BC1}
			&&g_{c}(t_{0})=\dot{g}_{c}(t_{0})=\ddot{g}_{c}(t_{0})=0,\\
			\label{BC2}
			&&g_{c}(t_{f})= \dot{g}_{c}(t_{f})=\ddot{g}_{c}(t_{f})=0,
		\end{eqnarray}
	\end{subequations}
	Furthermore, we can add more conditions depending on the problem to be solved. For the photonic GHZ, we require that the final cavity displacement be larger as we can at the final time $t_{f}$ the final cavity displacement be larger than we can. In this scenario, together with the boundary conditions in Eq.~(\ref{BC1}) and Eq.~(\ref{BC2}), we add the following constrains
	\begin{eqnarray}
		\label{condition_displacement}
		\alpha(t_{f})=-i\int_{0}^{t_{f}}g(s)e^{i\omega s}ds=d_{{\rm{max}}},
	\end{eqnarray} 
	corresponding to an arbitrary cavity displacement. On the other hand, for the generation of the GHZ state for qubits, apart from the boundary conditions given in Eq.~(\ref{BC1}) and Eq.~(\ref{BC2}) we require that 
	\begin{eqnarray}
		\label{condition_qubits}
		A_{n}(t_{f})=0, \quad B_{nn'}(t_{f})=\frac{\pi}{4}.
	\end{eqnarray} 
	In light of the experimental implementation of our model provided in Section~\ref{cQED}, the modulation of the longitudinal coupling strength $g(t)$ is through an adequate modulation of the external flux $\phi_{x}(t)$. Therefore, we propose a Fourier decomposition of the form $\phi_{x}(t)=\sum_{k}c_{k}\sin(\pi k ~t/t_{f})$ and plug this ans\"atze in Eq.~(\ref{equation_of_motion}) to compute the reversed engineering $g(t)$ (see Section~\ref{cQED} for the detailed derivation). In that way, the problem reduces to calculate the coefficient $c_{k}$ such $g_{c}(t)$ fulfills the boundary condition given in Eq.~(\ref{BC1}), Eq.~(\ref{BC2}) together with the Eq.~(\ref{condition_displacement}) for the photonic case, and Eqs.~(\ref{condition_qubits}) for the qubits, respectively.
	
	\begin{figure}[!t]
		\centering
		\includegraphics[width=1\linewidth]{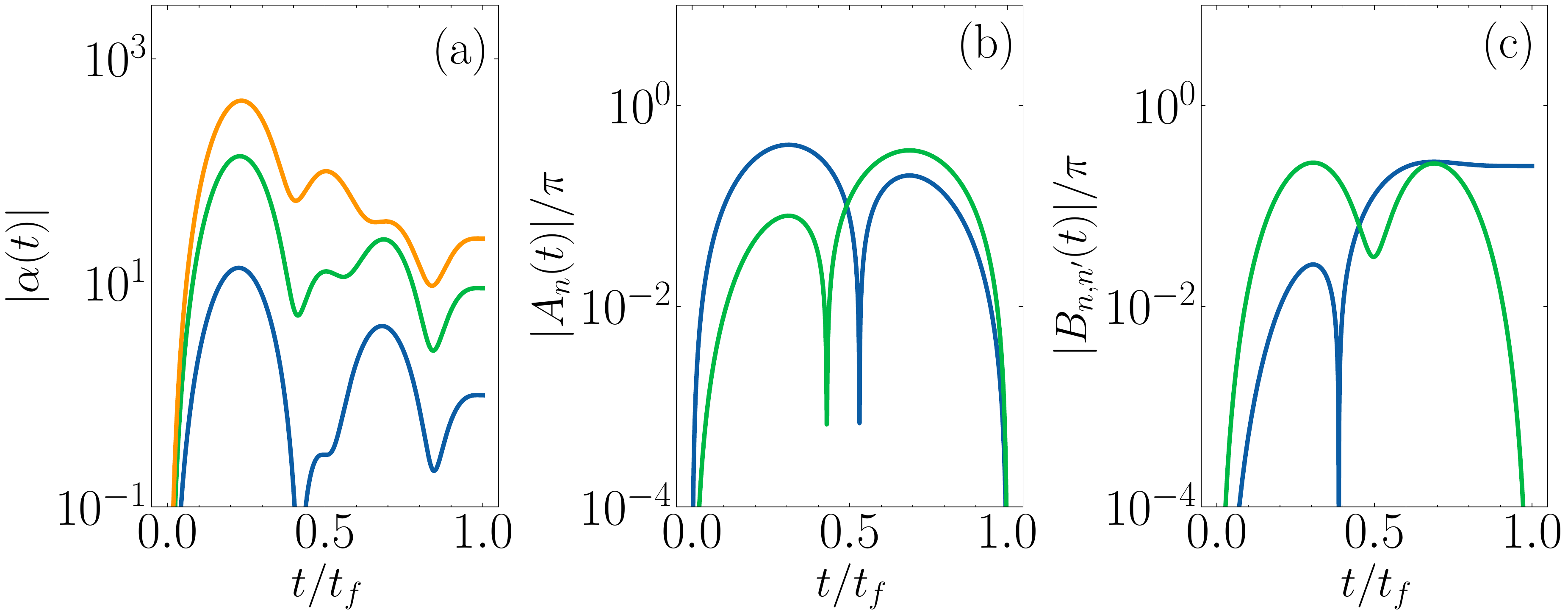}
		\caption{\textbf{Reversed enginnered modulation}. Reversely engineered (a) absolute cavity displacement $|\alpha(t)|$ as a function of the dimensionless time $t/t_{f}$, blue, orange and green lines correspond to the maximal cavity displacement of $d_{{\rm{max}}}=\{1,3,5\}$, respectively. (b) and (c) stand for the real (blue) and imaginary (green) parts of the reversely engineered $|A_{n}(t)|$ and $|B_{nn'}(t)|$ coefficients, respectively. The numerical simulations are performed with the same parameters as those in  Fig.~\ref{Fig_modulation}.}
		\label{Fig_conditions}
	\end{figure}
	
	We numerically calculate the coefficients $c_{k}$ using the gradient descent minimization package of python~\cite{Gradient_descen}. Fig.~\ref{Fig_modulation} shows the auxiliary variable $g_{c}(t)$ and the coupling strength $g(t)$ obtained through Eq.~(\ref{ELE}) as a function of the dimensionless time $t/t_{f}$ for three different maximal cavity displacement $d_{{\rm{max}}}=\{1,3,5\}$, respectively. It manifests that the maximal value of the coupling strength $g(t)$ strongly depends on the maximal value of the cavity displacement constraining the coherent state's size, unlike the qubit case, where we do not observe such behavior in the resulting $g_{c}(t)$ and $g(t)$, respectively. Moreover, it is shown that the modulations obtained in Fig.~\ref{Fig_modulation} satisfy the additional conditions stated in Eq.~(\ref{condition_displacement}) and Eqs.~(\ref{condition_qubits}) for the photonic and qubit case, respectively. We plot these quantities as a function of the dimensionless time $t/t_{f}$ depicted in Fig.~\ref{Fig_conditions}. {For the coupling strength $g/2\pi=21~{\rm{MHz}}$~\cite{PhysRevLett.115.203601} for the photonic system, we obtain a generation time $t_{f}=\pi/(40g)\equiv3.74~\rm{(ns)}$ achieving $\alpha(t_f)=d_{{\rm{max}}}$. Whereas, the qubit-based GHZ state can be generated within the timescale $t_{f}=\pi\omega/(80g^{2})\equiv1.89~\rm{(ns)}$ for a coupling strength $g/2\pi=114~{\rm{MHz}}$~\cite{PhysRevB.81.104524} where the conditions $A_{n}(t_f)=0$ and $B_{n,n'}(t_f)=\pi/4$ are fulfilled, respectively.}\par

	\subsection*{Multi-partite photonic state}
	\label{phGHZ}
	\begin{figure}[!t]
		\centering
		\includegraphics[width=.95\linewidth]{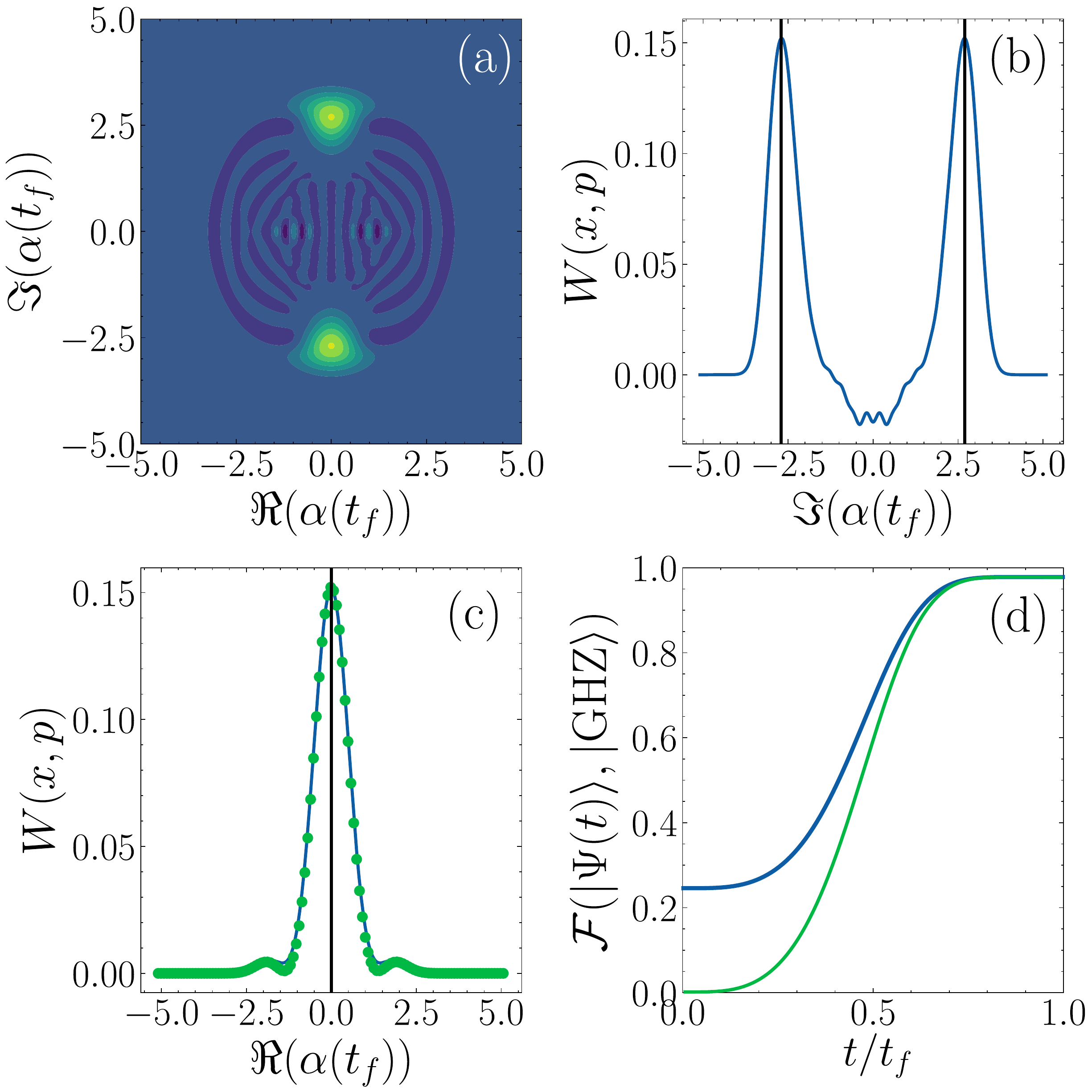}
		\caption{\textbf{Multi-partite photonic state} (a) Wigner representation of the reduced density matrix of the first field mode at final time $t_{f}$ for a system consisting into three field mode coupled to a single qubit. Projection of the Wigner function to the (b) coordinate space and (c) the momentum space, corresponding to a Gaussian function centered in $\pm d_{{\rm{max}}}=3$. (d) Fidelity $\mathcal{F}$ between the target Greenberger–Horne–Zeilinger state in Eq.~(\ref{GHZ_cat}) with the state $\ket{\Psi(t)}$ for two different maximal cavity displacement $d_{{\rm{max}}}=1$ to the blue line, and $d_{{\rm{max}}}=3$ for the green line, respectively. The numerical simulations are carried out with the same parameters as those in Fig.~\ref{Fig_modulation}.}
		\label{Fig_wigner_fidelity}
	\end{figure}
	
	We analyze the performance of the reversely engineered coupling strength $g(t)$ to generate photonic GHZ states for $M=3$ modes, see Hamiltonian of Eq.~(\ref{single_qubit}). In such a case, we assume that $g_{1}^{m}(t)\equiv g(t)$ $\forall m$. We quantify the Wigner function $W(x,p)$ of the reduced density matrix of the first field modes given by $\rho_{f_{1}}(t)={\rm{Tr}}_{\{f_{2},...,f_{M},q\}}[\ketbra{\Psi(t)}{\Psi(t)}]$, where the sub-indexes $f_{k}$, and $q$ corresponds to the partial trace over the $k$-th field mode and the qubit, respectively. Given that $W(x,p)$ for a coherent state $\ket{\alpha}$ corresponds to a Gaussian centered in $(\Re(\alpha),\Im(\alpha))$~\cite{Scully_quantum_optics_book}, at the final STA time $t_f$, where $\rho_{f_{1}}(t_{f}) = (1/2)(\ketbra{d_{{\rm{max}}}}{d_{{\rm{max}}}} + \ketbra{-d_{{\rm{max}}}}{-d_{{\rm{max}}}})$, the Wigner function represents two Gaussians centered at $\pm d_{{\rm{max}}}$. Fig.~\ref{Fig_wigner_fidelity}{(a)} shows the Wigner function on the phase space corresponding to two points far apart to each other from a distance $2d_{{\rm{max}}}$. Likewise, Fig.~\ref{Fig_wigner_fidelity}{(b)-(c)} shows the one-dimensional projection of $W(x,p)$ in the coordinate and momentum space, respectively. In concordance with Fig.~\ref{Fig_wigner_fidelity}, we appreciate that on the coordinate projection, we should expect to see a single Gaussian centered at zero (no imaginary component on the coherent state)  placed at $p=\pm d_{{\rm{max}}}$. In contrast, in the momentum projection, we have to see two overlapping Gaussian centered at $x=\pm d_{{\rm{max}}}$, respectively. From Fig.~\ref{Fig_wigner_fidelity}{(c)} we see a wide region where $W(x,p)$ takes negative values, showing the non-local nature of the multi-partite entangled state~\cite{Scully_quantum_optics_book}.
	
	Furthermore, we also calculate the Fidelity $\mathcal{F}(\rho,\sigma)=({\rm{Tr}}[\sqrt{\sqrt{\rho}\sigma\sqrt{\rho}}])^{2}$ ($\rho$ and $\sigma$ are arbitrary density matrices) between the state $\ket{\Psi(t)}$ obtained by numerically solve the Sch\"odinger equation of Eq.~(\ref{single_qubit}) with the target GHZ state defined in Eq.~(\ref{GHZ_cat}) for $M=3$ field modes for two different maximal field displacement $d_{{\rm{max}}}=\{1,3\}$ as depicted in Fig.~\ref{Fig_wigner_fidelity}(d). We observe that the reversed engineering protocol is insensitive to the final displacement $d_{{\rm{max}}}$, where we appreciate fidelities close to one for both modulations.\par
	\begin{figure}[!t]
		\centering
		\includegraphics[width=1\linewidth]{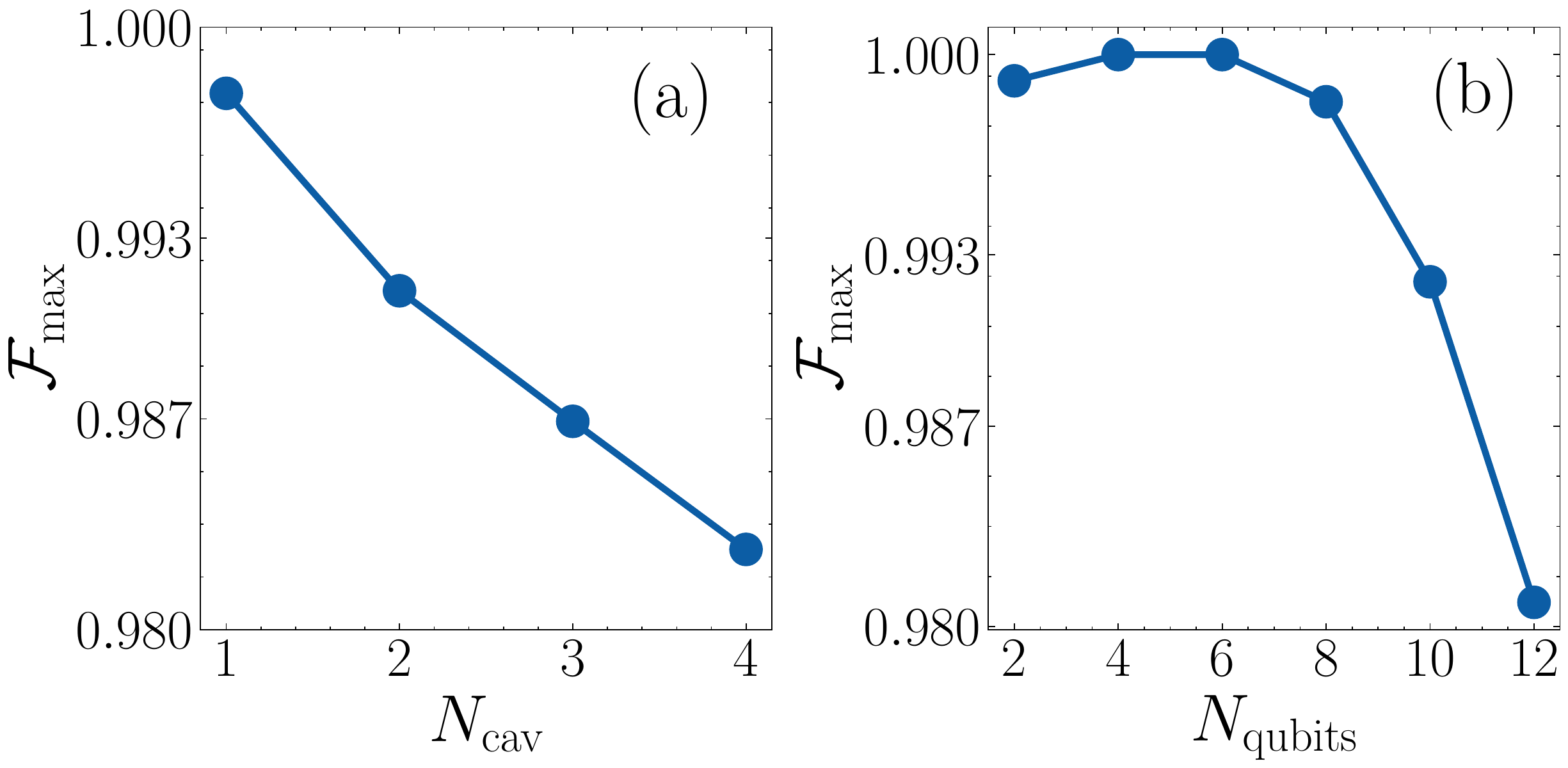}
		\caption{\textbf{Fidelity as function of the number of subsystems}. Fidelity $\mathcal{F}$ at the shortcut to adiabaticity time $t_{f}$ as a function of the number of system constituents for a system corresponding to (a) one qubit coupled to $M=\{1,2,3,4\}$ field modes with maximal cavity displacement $d_{\rm{max}}=3$, and (b) one field mode coupled to $N=\{2,4,6,8,10,12\}$, respectively. The numerical simulations are carried out with the same parameters as those in Fig.~\ref{Fig_modulation}.}
		\label{Fig_optimal_fidelity}
	\end{figure}
	Thus far, we have demonstrated that our reverse-engineering protocol provides a flexible way to engineer the longitudinal coupling strength that allows us to generate a multi-partite photonic states at a short timescale ($t_{f}\equiv3.2~\rm{(ns)}$), which is one order of magnitude shorter than the same protocol without the use of the reverse engineering approach ($T=47.61~\rm{(ns)}$). Moreover, our reverse engineering approach permit us fixing the final cavity displacement beyond the maximal distplacement $\alpha_{\rm{max}}=\pm2g_1^{m}/\omega_m$. Nevertheless, nothing is said about the scaling of the protocol in terms of the number of field modes. We should expect that the gating time $t_{f}$ does not scale from our calculations. To assure that, we solve the dynamics by increasing the number of field modes from $M=1$ to $M=4$ and compute the fidelity $\mathcal{F}(t_{f})$ at $t_{f}$ for each case. Those results are depicted in Fig.~\ref{Fig_optimal_fidelity}{(a)},  where we observe a high-fidelity generation of the cat Sch\"odinger states (the maximal fidelities obtained with the protocol are around $98\%-99\%$ for each case). Furthermore, we also see minor variations in the achieved maximal fidelity for the maximal number of modes simulated here. The exploration regarding a higher amount of field modes may be helpful to probe our conjecture. However, we are contained by the maximal displacement that bound the size of the whole Hilbert space, i.e., the maximal size of each field mode is chosen such $N_{{\rm{max}}}=|d_{\rm{max}}|+1$. In this sense, there exists a trade-off between the macroscopic size of the coherent state (related to the maximal displacement) with the maximal number of field modes to be simulated. 
	
	\subsection*{Multi-partite qubit state}\label{qbGHZ}
	\begin{figure}[!t]
		\centering
		\includegraphics[width=1\linewidth]{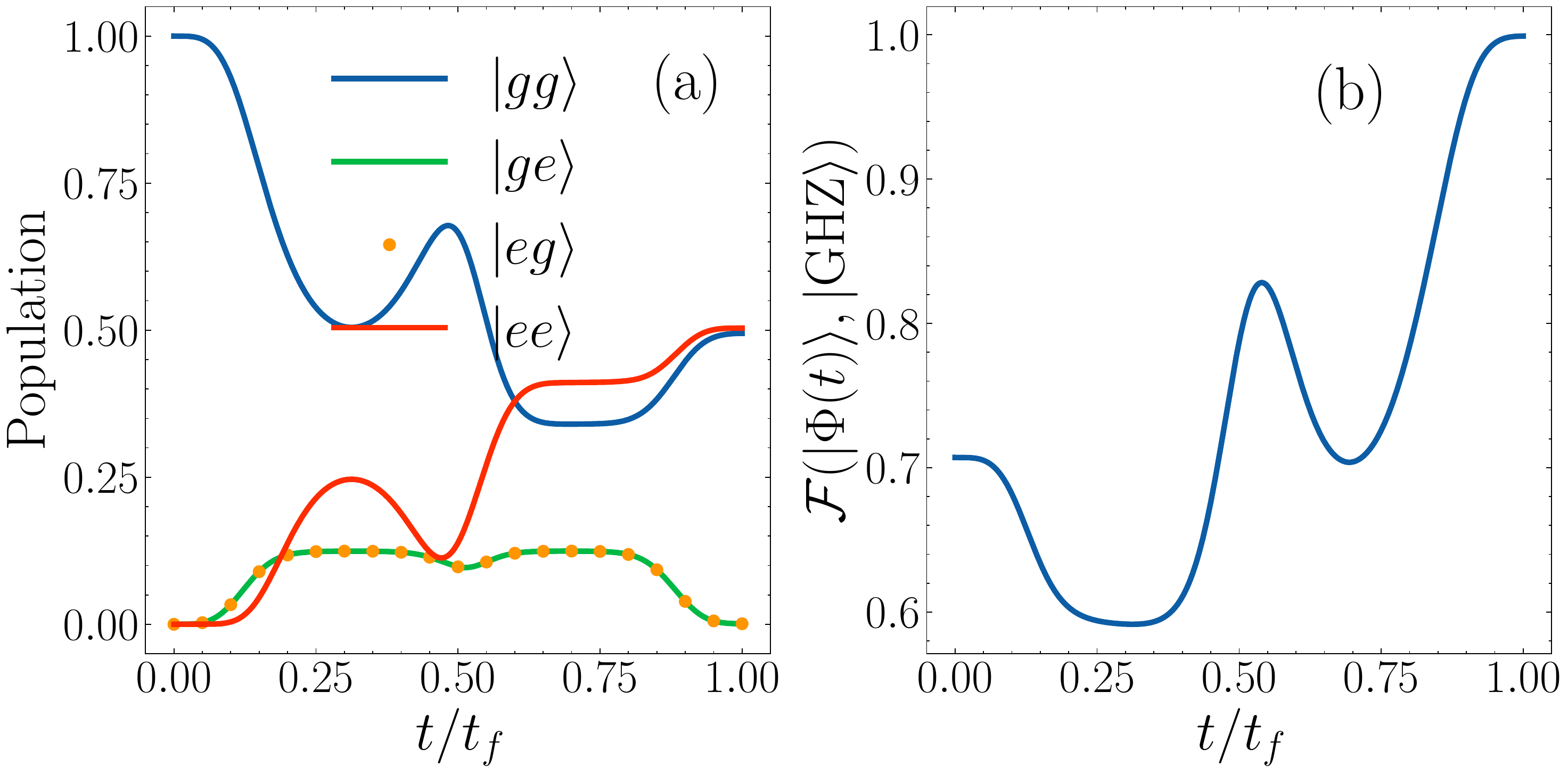}
		\caption{\textbf{Revsersed engineering multi-qubit state generation}. (a) Population evolution of the two-qubit states $\ket{gg}$ (continuous blue),  $\ket{ge}$ (continuous orange),  $\ket{eg}$ (dotted green), and  $\ket{ee}$ (continuous red) computed from the Hamiltonian in Eq.~(\ref{Hamiltonian_GHZ_qubit}) through the reverse- engineering protocol as a function of the dimensionless time $t/t_{f}$, we observe that at final time the state evolves to a GHZ state. (b) Fidelity $\mathcal{F}$ as a function of $t/t_{f}$ between the GHZ state of two qubits with the state $\ket{\Phi(t)}$. The numerical simulations are carried out with the same parameters as those in  Fig.~\ref{Fig_modulation}.}
		\label{Fig_pop_fid}
	\end{figure}
	
	Alternatively, we will analyze the performance of the reverse-engineering protocol to generate GHZ states in a qubit-based system. For doing so, we study the dynamics of the Hamiltonian in Eq.~(\ref{Hamiltonian_GHZ_qubit}) by considering $N=2$ qubits assuming homogeneous coupling strength between the qubits with the field mode, i.e., $g_{n}^{1}(t)=g(t)$ $\forall n$. Then, we can extend the result to a multi-qubit system up to 12 of them.

	Similar to the previous section, we look at a reduced part of the Hilbert space corresponding to the reduced density matrix of the $N$ qubits defined as $\rho_{q}(t)={\rm{Tr}}_{f}[\ketbra{\Phi(t)}{\Phi(t)}]$, where the sub-index $f$ corresponds to the trace over the field mode, where $\ket{\Phi(t)}$ is the solution of the Sch\"odinger equation for the Hamiltonian given in Eq.~(\ref{Hamiltonian_GHZ_qubit}). Fig.~\ref{Fig_pop_fid} {(a)} shows the population evolution for a system composing by $N=2$ qubits. We see that the occupation probabilities of the symmetric subspace $\{\ket{ge},\ket{eg}\}$ does not be equal to zero during all the dynamics because the terms related to their evolution in the time evolution operator given in Eq.~(\ref{U_GHZ_qubit_int_picture}) is not zero at all the times. In fact, the quantity $A_{n}(t)$ only vanishes at $t=0$, and $t=t_{f}$, respectively. Likewise, the population of the state $\ket{ee}$ keeps changing until $t_{f}$ achieves its optimal value. In Fig.~\ref{Fig_pop_fid}{(b)} we plot the fidelity $\mathcal{F}$ as a function of time with the target state given in Eq.~(\ref{GHZ_state_qubit}). Since the symmetric state contributes to the whole state's probability, fidelity tends to decrease until it reaches its maximal value around $\mathcal{F}=99\%$. Concerning the scaling of the generation time for the GHZ state, we obtain that the gating time does not scale with the number of qubits. In order to prove that, we calculate the fidelity $\mathcal{F}$ evaluated at $t=t_{f}$ by increasing the number of qubits from $N=\{2-12\}$. Fig.~\ref{Fig_optimal_fidelity}{(b)} shows the maximal fidelity as a function of the number of qubits. It is  observed that the fidelities $\mathcal{F}$ are bounded between $98\%-99\%$, thus demonstrating that the gating time does not scale with the number of qubits for achieving a large fidelity.\par

	Notice that in the derivation of the unitary transformation provided in Eq.~(\ref{SMGate}), no approximation has been made, which makes this protocol faster than other proposals based on dispersive interaction corresponding to second-order processes~\cite{PhysRevB.74.104503,PhysRevA.95.013845,Scirep.9.1380}. In this sense, the reverse-engineering protocol of generating these multi-parties entangled states allows us to accelerate its generation leading to generation times shorter than the tenth nanosecond scale, which mitigates the error produced by the unavoidable interaction with the environment. In the next section, we will quantify how the dissipation affects the performance of our protocols.
	
	On the other hand, more quantum control techniques are available such as  DRAG~\cite{DRAG1,DRAG2}, FAQUAD~\cite{FAQUAD1}, and GRAPE~\cite{GRAPE1,GRAPE2}. The latter corresponds to a gradient-based protocol that found the optimal modulation to control a system with decimated dynamics.\par
	The working principle of this algorithm relies upon assuming constant control fields for time windows. Under this approximation, we write the time-evolution operator without the time-ordered operator corresponding only to Hamiltonian exponentiation, and we update the control fields through their gradients. The complexity scales with a) the number of time steps of the control field, the Hamiltonian size, and its exponentiation. In our particular problem, a GRAPE-based implementation requires computing the exponentiation and their gradients for each time step at each minimization iteration, making them computationally challenging even for a modest number of cavities/qubits. Therefore, for the photonic GHZ, we must exponentiate matrices of dimension $2\otimes\dim(a)^M$, whereas, for the qubit multi-partite entangled state, the dimension scales as $2^N\otimes\dim(a)$. Contrary, the STA reversed engineering approach is less computationally demanding since we are only required to solve a set of differential equations that can be efficiently solved using standard numerical/analytical techniques.
	\par
	\subsection*{Dissipative dynamics}\label{dissipation}
	The next step is to analyze the performance of the reversed engineering approach to generate entangled states under dissipative and dephasing dynamics on qubits described by the following master equation
	\begin{eqnarray}
		\label{master_equation}\nonumber
		\frac{d\rho(t)}{dt}&=&-i[\mathcal{H}_{n}^{m},\rho(t)]+\sum_{m=1}^{M}\kappa_{m}\mathcal{L}[a_{m},\rho(t)]\\
		&+& \sum_{N=1}^{N}(\gamma_{n}\mathcal{L}[\sigma^{-}_{n},\rho(t)]+\gamma_{\phi,n}\mathcal{L}[\sigma^{z}_{n},\rho(t)]),
	\end{eqnarray}
	where $\mathcal{H}_{n}^{m}$ is the Hamiltonian given in Eq.~(\ref{model}), $\mathcal{L}[\mathcal{O},\rho(t)]=\mathcal{O}\rho(t)\mathcal{O}^{\dag}-\{\mathcal{O}^{\dag}\mathcal{O},\rho(t)\}/2$ is the Lindbladian operator describing the dynamics of a open quantum system~\cite{Breuer_book}. $\kappa_{m}$ and $\gamma_{n}$ stand for the relaxation ratio of the $m$-th cavity and the $n$-th qubits, respectively. Finally, $\gamma_{\phi,n}$ is the dephasing ratio of the $n$-th qubit.\par

	In superconducting circuits and cQED, energy relaxation appears as the consequence of the coupling between the quantum system with its electronic environment from the external circuitry, which we interpret as the interaction through an effective resistance. This noise profile is known as Nyquist noise or ohmic noise~\cite{ncomms12964,quintana2014}. Another source of relaxation relies upon quasiparticle decay~\cite{review,Serniak2018}, generated by the unpairing of Cooper-pairs due to electronic/thermal fluctuations. These electrons tunnel through the device, and we interpret them as an ohmic current leading to random spin flips. On the other hand, one of the dephasing sources corresponds to random spin flips in the superconducting metal forming the device due to the modulation of the external magnetic flux in the artificial atom~\cite{Koch2007}. Another source of error is produced by the longitudinal coupling with the quasiparticle bath that randomly changes the energy of the artificial atom~\cite{catelani2012,zanker2015}.\par
	In light of our experimental implementation that requires manipulating the external flux for the reversed engineering protocol, a way to mitigate errors associated with the tuning relies upon accessing to a flat energy spectrum in the artificial atom for achieving reduced flux sensitivity $\partial[\Omega_{n}]/\partial[\phi_{x}]\approx0$, which can be achieved with larger Josephson junctions~\cite{review}.\par
	Here, we will focus on the previous cases, where we consider a single-qubit interacting with $M$ modes described by $\mathcal{H}_{1}^{m}$ as in Eq.~(\ref{single_qubit}), and $N$ qubits interacting with a single cavity characterized by $\mathcal{H}_{n}^{1}$ in Eq.~(\ref{Hamiltonian_GHZ_qubit}). Without loss of generality, we assume homogeneous decay rates for the energy relaxation of the field mode, qubit relaxation, and dephasing rate, i.e., $\kappa_{m}=\kappa$, $\gamma_{n}=\gamma$, and $\gamma_{\phi,n}=\gamma_{\phi}$, respectively. We start by preparing the systems in the following state $\rho_{1}^{m}(0) = \ketbra{g}{g}\bigotimes_{m=1}^{M}\ketbra{0_m}{0_m}$ for the photonic case, and  $\rho_{n}^{1}(0) = \bigotimes_{n=1}^{N}\ketbra{g_{n}}{g_{n}}\otimes\ketbra{0}{0}$ for the pqubit based, respectively. We let them evolve until $t_{f}$, and compute the fidelity $\mathcal{F}$ at that time. For the GHZ cat state, we obtain that the fidelities for $M=1$ and $M=2$ is equal to $\mathcal{F}=\{0.998,0.996\}$. Due to the constraints imposed by the size of the Hilbert space, we cannot explore the system, including more field modes. For the qubit-based GHZ state, we obtain that the fidelities are given by $\mathcal{F}=\{0.999,0.998\}$ for $N=\{2,4\}$, respectively. Since the coherence times are longer than the generation time, we do not observe an appreciable change in the optimal fidelity. We have performed our numerical simulation considering the same physical parameters as Fig.~\ref{Fig_modulation}. For the decay rates we have chosen $\kappa/2\pi=1~{\rm{MHz}}$~\cite{PhysRevLett.115.203601} for the field modes, and $\gamma/2\pi=1/T_{1}$, with $T_{1}=40~{\rm{\mu s}}$ and $\gamma_{\phi}/2\pi=1/T_{2}$ with $T_{2}=40~{\rm{\mu s}}$~\cite{Wendin2017} for the two-level system, respectively.
\subsection*{Physical implementation}\label{cQED}
	\begin{figure}[!t]
		\centering
		\includegraphics[width=.85\linewidth]{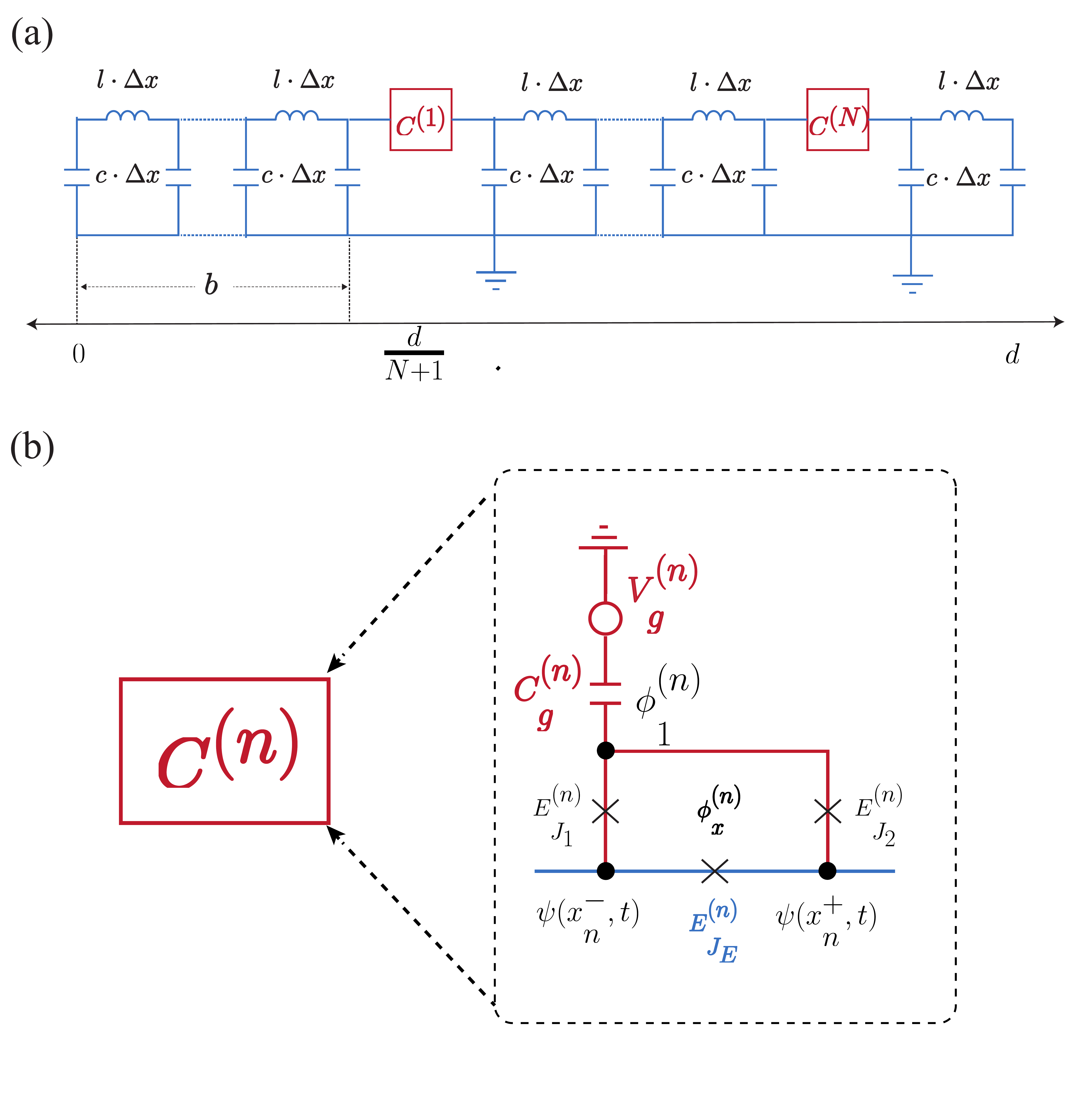}
		\caption{\textbf{Circuit quantum electrodynamic implementation}. Schematic illustration of our experimental proposal consisting of (a) a $\lambda/2$ coplanar waveguide resonator (CPWR) of length $d$ galvanically coupled to (b) $N$ artificial atoms $C^{(n)}$ formed by three Josephson junctions $E_{J_{\ell}}^{(n)}$ threaded by an external magnetic flux $\phi_{x}^{(n)}$. Moreover, we biased each artificial atom through a voltage source $V_{g}^{(n)}$ coupled through a capacitor $C_{g}^{(n)}$. We model the coplanar waveguide resonator as a series of LC circuits characterized by the capacitance and inductance per unit of length $c$ and $l$, respectively.}
		\label{Fig_circuit_implementation}
	\end{figure}	
	Finally, we propose an implementation for $\mathcal{H}_{n}^{m}$ in Eq.~(\ref{model}) based on cQED architecture consisting of a coplanar waveguide resonator (CPWR)~\cite{Goppl2008} of length $d$ with capacitance and inductance per unit length $l$ and $c$, respectively. We couple $N$ like-flux-qubit~\cite{Orlando1999,Mooij1999,Shcherbakova2015} artificial atom formed by two Josephson junctions (JJs) capacitances $\{C_{J_{1}}^{(n)}, C_{J_{2}}^{(n)}\}$ and Josephson energies $\{E_{J_{1}}^{(n)}, E_{J_{2}}^{(n)}\}$
to the CPWR through an embedded junction of capacitance $C_{J_{E}}^{(n)}$ and Josephson energy $E_{J_{E}}^{(n)}$ uniformly distributed along it~\cite{TLR_junction} (see Fig.~\ref{Fig_circuit_implementation}{(a)}). Moreover, we thread the artificial atom with an external magnetic flux $\varphi_{x}^{(n)}$ and driven by an external voltage $V_{g}^{(n)}$ through the capacitor $C_{g}^{(n)}$, as illustrated in Fig.~\ref{Fig_circuit_implementation}{(b)}. The circuit Lagrangian reads
\begin{eqnarray}
\label{lagrangian}
\mathcal{L} = \sum_{n=1}^{N}\big[\mathcal{L}_{\rm{q}}^{(n)} +\mathcal{L}_{\rm{int}}^{(n)}\big] + \mathcal{L}_{\rm{CPWR}},
\end{eqnarray}   
where
\begin{eqnarray}
\label{LLF1}\nonumber
\mathcal{L}_{\rm{q}}^{(n)} &=& \frac{C_{\Sigma_{1}}^{(n)}}{2}[\dot{\phi}_{1}^{(n)}]^2-q_{g}^{(n)}\dot{\phi}_{1}^{(n)}+E_{J_{1}}^{(n)}\cos\bigg(\frac{\phi_{1}^{(n)}}{\varphi_{0}}				\bigg)\\
&+&E_{J_{2}}^{(n)}\cos\bigg(\frac{\phi_{1}^{(n)}-\phi_{x}^{(n)}}{\varphi_{0}}\bigg),\\\nonumber
\label{LLint}
\mathcal{L}_{\rm{int}}^{(n)}&=&-C_{J_{1}}^{(n)}\dot{\phi}_{1}^{(n)}\dot{\psi}(x_{n}^{-},t)-C_{J_{2}}^{(n)}\dot{\phi}_{1}^{(n)}\dot{\psi}(x_{n}^{+},t)\\\nonumber
&-&C_{J_{2}}^{(n)}\dot{\phi}_{1}^{(n)}\dot{\phi}_{x}^{(n)}-C_{J_{2}}^{(n)}\dot{\psi}(x_{n}^{+},t)\dot{\phi}_{x}^{(n)}\\\nonumber
&+&E_{J_{1}}^{(n)}\sin\bigg(\frac{\phi_{1}^{(n)}}{\varphi_{0}}\bigg)\frac{\psi(x_{n}^{-},t)}{\varphi_{0}}\\
&+&E_{J_{2}}^{(n)}\sin\bigg(\frac{\phi_{1}^{(n)}-\phi_{x}^{(n)}}{\varphi_{0}}\bigg)\frac{\psi(x_{n}^{+},t)}{\varphi_{0}},\\\nonumber
\label{LLemb}
\mathcal{L}_{\rm{CPWR}}&=&\int_{0}^{d}\bigg[\frac{c}{2}[\partial_{t}\psi(x,t)]^2-\frac{1}{2l}[\partial_{x}\psi(x,t)]^2\bigg]dx\\\nonumber
&+&\sum_{n=1}^{N}\bigg[\frac{C_{\Sigma_{2}}^{(n)}}{2}[\partial_{t}\psi(x_{n}^{-},t)]^2 + \frac{C_{\Sigma_{3}}^{(n)}}{2}[\partial_{t}\psi(x_{n}^{+},t)]^2\\
&-&C_{J_{E}}^{(n)}\partial_{t}\psi(x_{n}^{-},t)\partial_{t}\psi(x_{n}^{+},t)\bigg].
\end{eqnarray}
Here, $\phi_{1}^{(n)}$ and $\psi(x,t)$ corresponds to the flux variables describing the artificial atom and the flux at the position $x$ of the CPWR, moreover, $C_{\Sigma_{1}}^{(n)}=C_{g}^{(n)}+C_{J_{1}}^{(n)}+C_{J_{2}}^{(n)}$, $C_{\Sigma_{2}}^{(n)}=C_{J_{1}}^{(n)}+C_{J_{E}}^{(n)}$ and $C_{\Sigma_{3}}^{(n)}=C_{J_{2}}^{(n)}+C_{J_{E}}^{(n)}$ are effective capacitances. Furthremore, $q_{g}^{(n)}=C_{g}^{(n)}V_{g}^{(n)}$ is the charge bias, $\varphi_{0}=\hbar/2e$ is the reduced quantum flux. Following the standard procedure of circuit quantization, and assuming that the junction forming the flux-qubit are identical and the embedded ones is different by a factor $\alpha^{(n)}$ we arrive a the following Hamiltonian
\begin{eqnarray}
\label{Model}\nonumber
\mathcal{H} &=& \sum_{m}^{M}  \omega_{m}a^{\dag}_{m}a_{m}+\sum_{n=1}^{N}\frac{\Omega_{n}}{2}\sigma_{n}^{z}\\
&+&\sum_{n,m} g_{n}^{m}(\varphi_{x}^{n})\sigma_{n}^{z}(a_{m}^{\dag}+a_{m}).
\end{eqnarray}
The detailed derivation of the above Hamiltonian can be found in the Supplementary Note 2. Here $a_{m}$ $(a^{\dag}_{m})$ correspond to the annihilation (creation) bosonic operator describing the $m$th mode of the CPWR with frequency $\omega_{m}$. Moreover, $\sigma^{z}_{n}$ is the $z$-component Pauli matrix describing the flux-qubit in the two-level approximation with transition frequency $\Omega_{m}$. Finally, $g_{n}^{m}(\varphi_{x}^{n})$ is the flux-dependent coupling strength between the CPWR with the $N$ two-level systems. Notice that we have neglected all the capacitive interactions since we are assuming the Josephson energy as the leading energy contribution in the artificial atom. Likewise, due to the presence of the embedded junction in the CPWR, we obtain an antisymmetric spatial profile in the current/voltage of the resonator. This eliminates the coupling between the artificial atom and the node $\psi(x_{n}^{+},t)$ of the CPWR, leading to a longitudinal interaction between both subsystems.

\section*{Conclusion}\label{conclusion}

In this article, we have proposed the reverse-engineering method to design a longitudinal interaction to generate multi-partite entangled states in photonic and qubit-based systems. 
Such approach suggests a modulation that satisfies the desired criteria under suitable conditions. In our case, we constrain the modulation to achieve a fixed cavity displacement to generate photonic multi-partite entangled states. In contrast, we impose the vanishing of some unwanted interaction to implement a S\o rensen-M\o lmer quantum gate to generate  multi-partite qubit states, respectively. As a result, we obtain a fast generation time (less than tens of nanoseconds) with the current state-of-the-art superconducting quantum circuits architecture. Notably, the gating time does not scale with the number of either field modes or qubits, allowing us to generate entangled quantum states containing many parties. Since the generation time is shorter than the coherence time of the subsystem, the unavoidable effect of the environment does not induce a significantly detrimental impact on the final fidelity generation. The renewed attention to the protocols provided by STA in the context of cQED may pave the way to accelerate and improve several quantum tasks by engineering the adequate pulses sequences and modulation that are fundamentally constrained in the quantum platform. 
	\section*{ACKNOWLEDGMENTS}
	F. A. C. L thanks to A. Parra-Rodriguez, for helpful discussion. This work is supported by NSFC (12075145), STCSM (Grants No. 2019SHZDZX01-ZX04),  EU FET Open Grant EPIQUS (899368), HORIZON-CL4-2022-QUANTUM-01-SGA project 101113946 OpenSuperQPlus100 of the EU Flagship on Quantum Technologies, the Basque Government through Grant No. IT1470-22, the project grant PID2021-126273NB-I00 funded by MCIN/AEI/10.13039/501100011033 and by ``ERDFA way of making Europe" and  ``ERDF Invest in your Future",  Nanoscale NMR and complex systems (PID2021-126694NB-C21) and QUANTEK project (Grant No. KK-2021/00070). 
	 F. A. C. L. thanks to the German Ministry for Education and Research, under QSolid, Grant no. 13N16149. J. C. R acknowledges to Financiamiento Basal para Centros Científicos y Tecnológicos de Excelencia (Grant No. AFB220001). X. C. acknowledges ayudas para contratos Ramón y Cajal--2015-2020 (RYC-2017-22482).
	\section*{COMPETING INTERESTS}
	The authors declare no competing interests.
	\section*{DATA AVAILABILITY}
	The data that support the findings of this study are available from the corresponding author, F. A. C. L, upon reasonable request.
	\section*{AUTHOR CONTRIBUTIONS}
	All authors originate the project. F. A. C. L. carried out the numerical simulations in collaboration with JC. R. X.C. provides the methods of STA. All authors contributed to verifying the results and writing the manuscript.

\end{document}